# Direct solutions for normal depths in curved irrigation canals

X.Y.Zhang*, L.Wu

*College of Water Resources and Architectural Engineering, Northwest A&F University, Yangling 712100, P.R.China*

A B S T R A C T

The normal depth is an important hydraulic element for canal design, operation and management. Curved irrigation canals including parabola, U-shaped and catenary canals have excellent hydraulic performance and strong ability of anti-frost heave, while the normal depths in the governing equations of the current common methods have no explicit analytical solution. They are only indirect methods by using trial procedures, numerical methods, and graphical tools. This study presents new direct formulas for normal depth in curved irrigation canals by applying for Marquardt method. The maximum relative error of the proposed formulas is less than 1% within the practice range by comparative analysis, and they are simple and convenient for manual calculations. The results may provide the reliable theoretical basis and useful reference for the design and operation management of irrigation canals.

## 1. Introduction

Compared with trapezoidal and rectangular canals, the stress condition of the curved canals including parabola, U-shaped and catenary canals has been improved [1]. They are beneficial to the prevention of frost heaving of soil matrix and have excellent hydraulic performance. It is a favorable choice in irrigation works, especially in cold regions. But the governing equation for the normal depth is implicit and no analytical solutions in design and operation management of curved irrigation canals. There are some indirect methods to solve these equations now, for example, the trial and error procedure, graphical tools, iteration procedure, and so on. But these methods have their disadvantages, as follows, (1) the trial and error procedure is time consuming and difficult to control; (2) the descriptive geometry solution is susceptible to thewrong reading and interpolation error; (3) the iteration process is easy to control the precision and error, but is susceptible to the iterative format and iterative initial value and always causes iteration non-convergence, or more iteration times and so on. So they can't be developed and put into the practice. This results in a large number of studies on the normal depth of open channels in the world [2,3]. The approximation calculation formulas of the normal depth are available for a half-cubic parabola canal [4], U-shaped canal [5,6], trapezoid canal and round canal[7], but not for catenary canal.

In this study, direct solutions for the normal depths are presented for curved canals including parabola, U-shaped and catenary canal by Marquardt method based on the NLIN procedure using SAS software. The proposed formulas either fill the current gap or improve upon existing formulas in terms of accuracy and simplicity.

*Corresponding author.
*Email address*:xnzxy@163.com(X.Y.Zhang)

| Notations | |
|---|---|
| $x$ | Transverse coordinates |
| $y$ | Longitudinal coordinates |
| $K$ | Parabola shape parameter |
| $A$ | Cross section area |
| $B$ | Width of the canal at the water surface |
| $h_0$ | Normal flow depth |
| $P$ | Wetted perimeter |
| $r$ | Bottom arc radius |
| $a$ | height of bottom arc |
| $2\theta$ | Central angle of section bottom arc |
| $m$ | Side slope coefficient |
| $b$ | Catenary shape parameter |
| $cc$ | constant |
| $Q$ | Discharge |
| $S_0$ | Canal bottom slope |
| $n$ | Roughness coefficient |
| $Q_t$ | Transitional flow rate |
| $SS\%$ | Relative error |
| $\eta_p = \sqrt{Kh_0}$ | Characteristic depth(parabola cross section) |
| $\varepsilon_p = \left(nQ/\sqrt{S_0}\right)^{\frac{3}{5}} K^{\frac{8}{5}}$ | Section characteristic parameter(parabola cross section) |
| $\eta_u = \dfrac{h_0}{r}$ | Dimensionless relative normal depth(U-shaped cross section) |
| $\varepsilon_u = \left(nQ/\sqrt{S_0}\right)^{0.6} / r^{1.6}$ | Section characteristic parameter(U-shaped cross section) |
| $\lambda = 1/\sqrt{1+m^2}$ | Dimensionless parameter(U-shaped cross section) |
| $\eta_c = \dfrac{B}{2b}$ | Dimensionless parameter(catenary cross section) |
| $\varepsilon_c = b^{-1.6}\left(nQ/\sqrt{S_0}\right)^{0.6}$ | Section characteristic parameter(catenary cross section) |

*Subscripts*

| | |
|---|---|
| $p$ | Denotes parabola cross section |
| $u$ | Denotes U-shaped cross section |
| $c$ | Denotes catenary cross section |

A comparison of the accuracy of the proposed and existing solutions is also presented. And this study may provide the reliable theoretical basis and useful reference for the design and operation management of irrigation canals.

## 2. Geometrical properties of the canal cross-sections

*2.1 Parabola canal*

Considering the coordinate system of Fig. 1, the parabola equation can be expressed as follows.

$$y = Kx^2 \qquad (1)$$

where $x$, $y$ are the transverse and longitudinal coordinates, respectively. $K$ is parabola shape parameter.

Then the geometric elements of the parabola canal section can be written as,

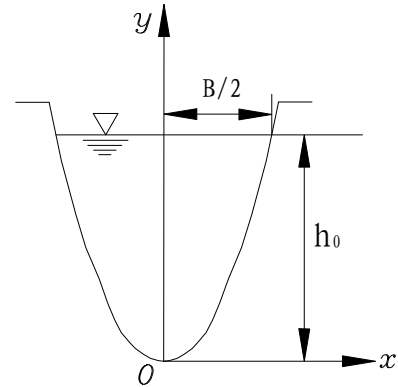

Fig.1 Cross section of parabolic canal

$$\left. \begin{aligned} A &= \dfrac{2}{3}Bh_0 \\ P &= \dfrac{B}{2}\sqrt{1+K^2B^2} + \dfrac{1}{2K}\ln\left(KB+\sqrt{1+K^2B^2}\right) \end{aligned} \right\} \quad (2)$$

where $A$ is cross-section area, $P$ is wetted perimeter; $B$ is width of the canal at the water surface; $h_0$ is flow depth.

*2.2 U-shaped canal*

Fig. 2 shows a U-shaped canal section, the applicable equations are defined as the piecewise functions changing with flow depth because the water-carrying section is consisted of bottom arc and upper line segment.

when $h_0 \leq a$,

$$A = r^2 \arccos\left(1 - \frac{h_0}{r}\right) - (r - h_0)\sqrt{r^2 - (r - h_0)^2}$$
$$P = 2r \arccos\left(1 - \frac{h_0}{r}\right)$$
(3)

when $h_0 > a$,

$$A = (h_0 - a)\left[\frac{2r}{\sqrt{1+m^2}} + (h_0 - a)m\right] + r^2\left(\theta - \frac{m}{1+m^2}\right)$$
$$P = 2\left[r\theta + (h_0 - a)\sqrt{1+m^2}\right]$$
(4)

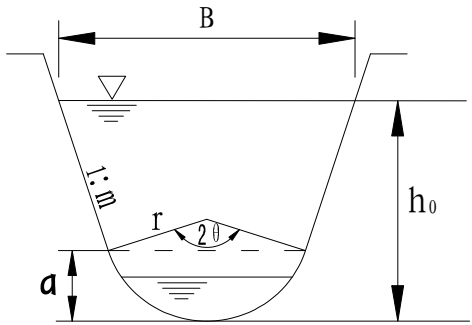

Fig.2 Cross section of U-shaped canal

where $r$ is bottom arc radius, $a$ is height of bottom arc, $a = r(1 - \cos\theta)$, $2\theta$ is central angle of section bottom arc, $m$ is side slope coefficient of sidewall line segment.

*2.3 Catenary canal*

Considering the coordinate system of Fig. 3, the catenary equation is presented as follows.

$$y = b \cdot ch\frac{x}{b} = \frac{b}{2}\left(e^{\frac{x}{b}} + e^{-\frac{x}{b}}\right) \qquad b > 0 \qquad (5)$$

where $b$ is catenary shape parameter. Then, the applicable equations can be expressed as,

$$h_0 = b\left(ch\frac{B}{2b} - 1\right)$$
$$A = B \cdot a \cdot ch\frac{B}{2b} - 2b^2 \cdot sh\frac{B}{2b}$$
$$P = 2a \cdot sh\frac{B}{2b}$$
(6)

The meanings of symbolic in the equation are as the previous.

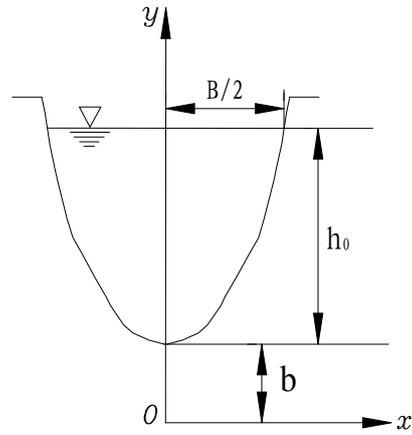

Fig.3 Cross section of catenary canal

## 3. methodology

Basic equation of open channel uniform flow is fitted gradually to find the optimal model parameters using Marquardt method in SAS software. Marquardt method is an effective method to solve the question of fitting non-linear equation and one of the most popular method in the field of solving non-linear equation at present [8-13]. It is to assess whether the fitting formula reaches the best fitting result by using the iterative program to calculate the residual sum of squares. The iterative procedure is end and the fitting formula is the best one when the minimum residual sum of squares exists. Marquardt method has the characteristics of well adaptable to formula, low requirement to iterative initial value and easy to convergence, etc.

## 4. Computation of normal depth

The movement principle of uniform flow in open channel is the fundamental basis for hydraulic calculation of open channel. In the engineering practice, theory of uniform flow in open channel is not only an important basis for channel design and management, but also the foundation of non-uniform flow theory in open channel. Its basic equation is as follows,

$$Q=\frac{\sqrt{S_0}}{n}\cdot\frac{A^{5/3}}{P^{2/3}} \qquad (7)$$

where $Q$ is discharge, $S_0$ is channel bottom slope, $n$ is roughness coefficient. It is true to the region of quadratic resistance law.

### 4.1 parabola canal

Substituting for $A$ and $P$ from Eq.(2) into Eq.(7) yields the following form,

$$cc\cdot\varepsilon_p=\frac{\eta_p^3}{\left[2\eta_p\sqrt{1+4\eta_p^2}+\ln\left(2\eta_p+\sqrt{1+4\eta_p^2}\right)\right]^{\frac{2}{5}}} \qquad (8)$$

where $\eta_p$ is characteristic depth of parabola canal with $\eta_p=\sqrt{Kh_0}$, $\varepsilon_p$ is characteristic section parameter with $\varepsilon_p=\left(\frac{nQ}{\sqrt{S_0}}\right)^{\frac{3}{5}}K^{\frac{8}{5}}$, $cc$ is a constant with $cc=0.5684$.

The solution of the normal depth is to get the characteristic depth $\eta_p$ in Eq. (8). But Eq. (8) is the complex implicit function of $\eta_p$ with no analytic solution. So we optimize and fit Eq. (8) based on the nonlinear regression procedure (NLIN) using SAS software. At first, a series of the non-linear models and their corresponding parameters are supposed. Then the direct calculation formula of $\eta_p$ has derived with aimed at the minimum sum of deviation square using Marquardt method.

$$\left.\begin{array}{l}\eta_p=e^{-0.016+0.075\sqrt{\varepsilon_p}+0.383\ln\varepsilon_p} \quad \varepsilon_p\leq 1 \\ \eta_p=0.055+0.99\varepsilon_p^{5/11} \quad \varepsilon_p>1\end{array}\right\} \qquad (9)$$

The calculation formula of the relative error can be expressed as,

$$SS\%=(\eta_{p^*}-\eta_p)/\eta_p\times 100\% \qquad (10)$$

where $SS\%$ is the relative error, $\eta_{p^*}$ is the calculation value of characteristic depth from the formula (9), $\eta_p$ is given value randomly in the engineering practice.

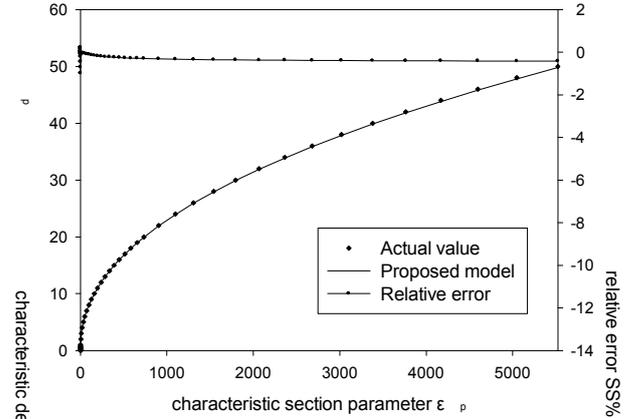

Fig.4. Comparison of actual and proposed characteristic depth in parabolic canal

The absolute value of relative error is less than 1%. And the minimum is 0.0057%. The relative error distribution is shown in Fig. 4. As noted, the proposed solution formula (9) has the high precision, simple formula form, clear physical concept and convenient calculation.

### 4.2 U-shaped canal

Substituting for $A$ and $P$ from Eqs.(3) and (4) into Eq. (7) yields the uniform flow basic equation of U-shaped canal. Because the hydraulic calculation of U-shaped canals changes with the flow depth, the flow depth needs to be judged according to actual flow rate $Q$ at first. At present, the actual flow rate in canal is used to be the transitional flow rate when the flow depth is exactly equal to the bottom arc height. It can be expressed by $Q_t$ as follows,

$$Q_t=\frac{\sqrt{S_0}}{n}\left[\frac{r^8\left(\arcsin\left(\frac{1}{1+m^2}\right)-\frac{m}{1+m^2}\right)^5}{4\arcsin\left(\frac{1}{1+m^2}\right)^2}\right]^{1/3} \qquad (11)$$

When the actual flow rate $Q$ in canal is less than the transitional flow rate $Q_t$, the flow depth $h_0$ is less than the bottom arch height $a$; when the flow rate $Q$ is greater than the transitional flow rate $Q_t$, the flow depth $h_0$ is greater than the bottom arc height $a$. That is to say, when $0<Q\leq Q_t$, then $0<h_0\leq a$,

$$cc \cdot \varepsilon_u = \frac{\arccos(1-\eta_u) - (1-\eta_u)\sqrt{1-(1-\eta_u)^2}}{[\arccos(1-\eta_u)]^{0.4}} \quad (12)$$

When $Q>Q_t$, then $h_0>a$,

$$cc \cdot \varepsilon_u = \frac{\arcsin\lambda + (\eta_u + \sqrt{1-\lambda^2} - 1)\left[\lambda + \sqrt{1/\lambda^2 - 1}(\eta_u + \sqrt{1-\lambda^2} - 1)\right]}{\left[\arcsin\lambda + 1/\lambda(\eta_u + \sqrt{1-\lambda^2} - 1)\right]^{0.4}}$$

(13)

where $\eta_u$ is the dimensionless relative normal depth with $\eta_u = \frac{h_0}{r}$, $\varepsilon_u$ is the section characteristic parameter with $\varepsilon_u = \left(\frac{nQ}{\sqrt{S_0}}\right)^{0.6} / r^{1.6}$, $\lambda$ is the dimensionless parameter with $\lambda = \frac{1}{\sqrt{1+m^2}}$, cc is a constant with $cc = 2^{0.4}$.

To get the normal depth is to solve formulae (12) and (13). But they include transcendental equation of anti-trigonometric function with no analytic solutions.

In practice, when $0<Q\leq Q_t$, $h_0 \leq a = r(1-\cos\theta)$, the central angle of section bottom arc $\theta \in (0,\pi]$, the value range of $\eta_u$ is $\eta_u \in (0,1]$. If putting $\eta_u$ value into formula (12), the span of $\varepsilon_u \in (0,1.0]$ is available. The side slope coefficient range of U-shaped canal side wall is $0 \leq m \leq 4$ in practical engineering. When $m=0$, the upside of the channel is rectangle and $\lambda = 1$. And when $m \neq 0, \lambda \neq 1$ is available. When $Q>Q_t$, $h_0 > a = r(1-\cos\theta)$, $\eta_u = \frac{h_0}{r} > 1-\cos\theta = 1 - \frac{m}{\sqrt{1+m^2}}$ is available. Substituting for $\lambda$ and $\eta_u$ into Eq. (13) yields $\varepsilon_u \in [1.0, 629.14]$.

Likewise, the optimal fitting models are available by the nonlinear regression in SAS software on Equations (12) and (13). That is, the direct hydraulic calculation formulae of normal depth in U-shaped canal are as follows,

$$\eta_u = -0.0019 + 0.14\varepsilon_u^{0.5} + 1.425\varepsilon_u - 1.776\varepsilon_u^{1.5} + 2.361\varepsilon_u^2 - 1.718\varepsilon_u^{2.5} + 0.577\varepsilon_u^3$$

$$\varepsilon_u \in (0, 1.0] \quad (14)$$

$$\left.\begin{array}{l}\eta_u = -0.57 + 1.587\varepsilon_u^{5/3} \quad \lambda = 1 \\ \eta_u = e^{-0.1927 + 0.6265\ln\varepsilon_u + 0.898\lambda^{1.5} - 0.0635/\lambda} \quad \lambda \neq 1\end{array}\right\}$$

$$\varepsilon_u \in [1.0, 629.14] \quad (15)$$

The precision evaluation on the formulae (14) and (15) are conducted through the error analysis. Firstly, the transitional flow rate $Q_t$ is determined in terms of the hydraulic elements of canal section, and the actual flow rate is compared with the transitional flow rate. When the actual flow rate $Q$ is less than the transitional flow rate $Q_t$, the flow depth in canal is below the bottom arc height and Eq. (12) is employed; when the actual flow rate is greater than the transitional flow rate, the flow depth is up to the bottom arc height and Eq. (13) is employed. For any given $\lambda$ and $\eta_u$ value in their range, $\varepsilon_u$ value is decided by the Eq. (12) or (13). Secondly, substituting for $\lambda$ and $\varepsilon_u$ value into the formula (14) or (15) yields a new $\eta_u$ value (denoted by $\eta_{u*}$). The calculation formula of relative error for the dimensionless relative normal depth is the same as the formula (10). The analysis results are shown in Fig. 5.

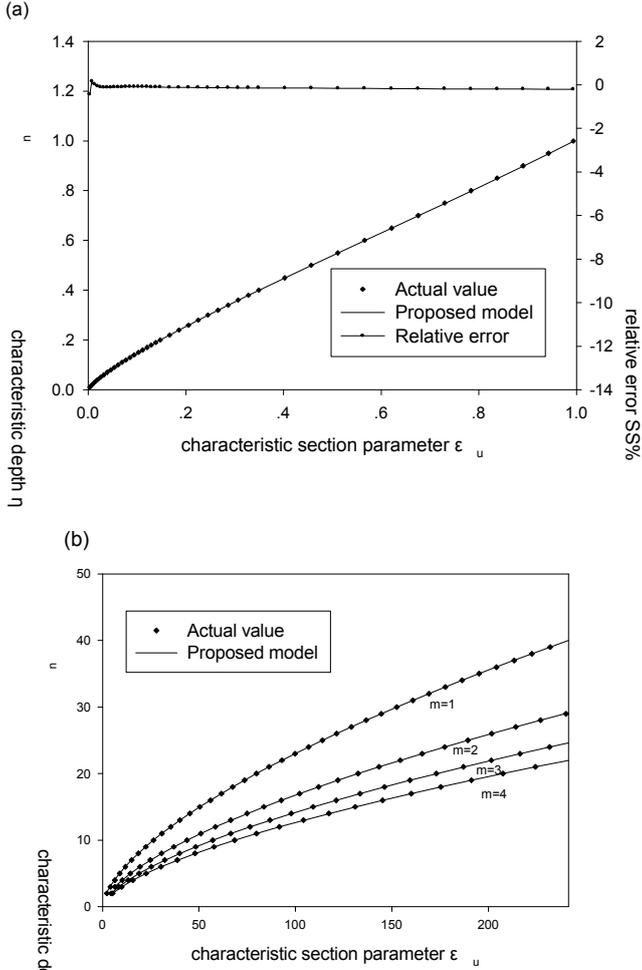

Fig. 5 Comparison of actual and proposed characteristic depth in U-shaped canal
(a) $h_0 \leq a$; (b) $h_0 > a$

Fig. 5 shows that the absolute value of relative error of direct calculation formulae (14) and (15) are less than 1%. When $h \leq a$, the absolute value of relative error is always less than 0.44%; when $h > a$ and $m=0$, the relative error is less than 0.08%; when $m \neq 0$, the error increases, but its absolute value is below 1%. We will get to know the proposed formulae have a higher precision. Moreover, they have simple formula form, clear physical concept, and strong universality.

### 4.3 Catenary canal

Substituting Eq. (6) into Eq. (7), the uniform flow basic equation of catenary canal is available as follows,

$$cc \cdot \varepsilon_c = \frac{\eta_c \cdot ch\eta_c - sh\eta_c}{(sh\eta_c)^{0.4}} \quad (16)$$

where $\eta_c$ is dimensionless parameter with $\eta_c = \frac{B}{2b}$; $\varepsilon_c$ is canal section characteristic parameter with $\varepsilon_c = b^{-1.6}\left(\frac{nQ}{\sqrt{S_0}}\right)^{0.6}$, cc is a constant with $cc = 2^{-0.6}$.

Eq. (16) is a transcendental equation including hyperbolic function with no analytic solutions and direct calculation.

The data range of dimensionless parameter $\eta_c$ is determined at first, and there is no restriction for the data range of dimensionless parameter $\eta_c$ at present. But the optimal hydraulic section is the design basic in the hydraulic design of irrigation canals. The condition of optimal hydraulic section for catenary canal is $\eta_m = \frac{B}{2b} = 1.6061$. So it is expanded by 30 times on the basis of this relation. That is to say, the data range of $\eta_c$ is given by $\eta_c \in [1.6061/30, 1.6061 \times 30] = [0.0535, 48.183]$. This range satisfies the practical engineering needs adequately. Substituting $\eta_c$ into Eq. (16), the range of $\varepsilon_c \in [0.000165, 1.1182E+14]$ is available.

In the same way, the direct calculation formula of normal depth for catenary canal is gotten with the non-linear regression model in SAS software.

$$\begin{cases} \eta_c = 0.056 - 0.0079\varepsilon_c - 0.14\sqrt{\varepsilon_c}\ln\varepsilon_c + 1.457\sqrt{\varepsilon_c} + 0.0048\ln\varepsilon_c \\ \quad \varepsilon_c \in [0.00025, 3.529] \\ \eta_c = -4.85 + 1.6\ln\varepsilon_c + 12.8/\ln(2^{-0.6}\varepsilon_c) - 28.95/\varepsilon_c \\ \quad \varepsilon_c \in [3.529, 1.6949E+14] \end{cases} \quad (17)$$

The relative error is also expressed by the formula (10). Fig. 6 shows that the absolute value of the maximum relative error for the formula (17) is less than 0.93%. That is to say that the proposed formula (17) has a higher precision. Likewise, it has simple formula form, clear physical concept, and strong universality.

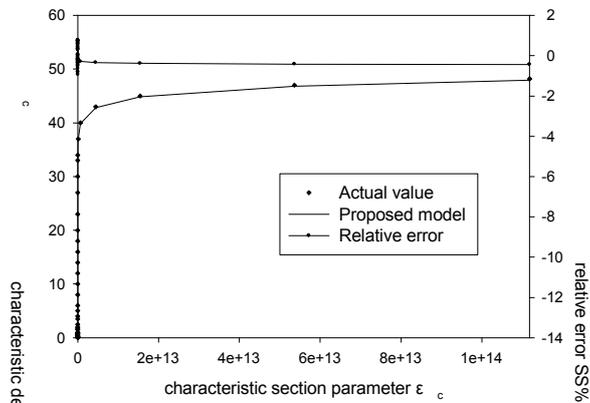

Fig.6. Comparison of actual and proposed characteristic depth in catenary canal

## 5. Conclusions

The curved irrigation canals have uniform cross-section structure and excellent hydraulic performance. The normal depth is an important parameter in the design, operation, and maintenance of irrigation canals. But their governing equations are complicated and cannot be directly solved for curved cross-section. The normal depth solving is traditionally performed using trial procedures, numerical methods, and graphical tools. These methods have the disadvantages such as complex, low accuracy, not easy to use, etc.

Based on the principle of open channel uniform flow, the normal depth direct hydraulic calculation formulae for parabola, U-shaped and catenary canal were established using Marquardt method in this study.

The absolute value of relative errors of the proposed formulae is less than 1% within the practical scope of the project without exception. It is less than 1% for parabola canal in which the minimum is only 0.0057% and less than 0.93% for catenary canal. For U-shaped canal, the absolute value of relative error is less than 0.44% when h≤a and 0.08% when h>a and m=0. It is thus obvious that the proposed formulae have high accuracy. Moreover, the proposed formulae, with the simple form and the clear physical concept, exhibit both simplicity and easy-to-use. This promotes convenience and precision of the design and management of irrigation canals. Therefore, the efficient computational tools presented in this paper will be useful in the design and management of irrigation canals.